\documentclass{aa}
\usepackage{graphicx}
\usepackage{txfonts}

\begin{document}

\title{Truncated stellar disks in the near infrared. I. Observations}
\titlerunning{Truncated disks} 

\author{E. Florido\inst{1}, E. Battaner\inst{1}, A. Guijarro\inst{1,2}, F. Garz\'on\inst{3,4} \and A. Castillo-Morales\inst{1}}

\offprints{E. Florido (estrella@ugr.es)}

\institute{Departamento de F\'{\i}sica Te\'orica y del Cosmos, Universidad de Granada, Spain.
   \and
      Centro Astron\'omico Hispano Alem\'an, Almer\'{\i}a, Spain 
   \and
      Instituto de Astrof\'{\i}sica de Canarias, 38200 La Laguna, Spain
   \and
      Departamento de Astrof\'{\i}sica, Universidad de La Laguna, Tenerife, Spain }

   \date{}

\abstract{We present NIR surface photometry of 11 edge-on galaxies obtained in the course of a long term project aimed at analysing the occurrence and type of the truncation of the outer disks. Observations were carried out at the 1.5 m CST (Carlos S\'anchez Telescope) in Tenerife (Spain) using the CAIN infrared camera. 7 galaxies exhibit clear truncation on their disk profiles and 4 galaxies were observed to be clearly untruncated within observational limits. We describe the truncations as real, smooth and complete (as suggested by extrapolation and in the sense that the measured truncation curve goes into the noise at a truncation radius $R_{tr}$), following a decline proportional to $(R_{tr} -R)^{-n}$  (where $R$ is the radius). Despite its deep photometric reach, the data presented do not permit a detailed exploration of the region where optical data show a second slope. Special care  was taken concerning the surface brightness deprojection of edge-on galaxies, which was carried out by two methods, one comprising the inversion of Abel's integral equation and the other following a numerical method. These methods gave nearly identical results. NIR observations of truncations could differ from observations in the optical, since the two domains trace different stellar populations.
}

\maketitle

\section{Introduction}

It is widely accepted that the galactic disk brightness profiles, at least in the inner regions where the contribution of the bulge to the observed emission is still low, can be reproduced with an exponential function. At large radii, it is found that the stellar distribution often departs from this simple description and decreases much faster. One of the goals of this paper is to contribute to determining the functional form of such a rapid decline.

This decline of the stellar density was discovered by van der Kruit (1979), who denoted it as truncation. In this and subsequent papers (van der Kruit \& Searle, 1981a, 1981b) the basic properties of truncations were described as well as the first theoretical ideas to explain them. Some recent studies have found that there is no complete truncation, but rather a sharp change in the exponential constant, called a ``break'' (Pohlen et al. 2002a, Pohlen et al. 2004 and references therein). 

Let us define the truncation curve as
\begin{equation}
   \tau(R) = \mu(R) - \mu_D(R)
\end{equation}
as in Florido et al. (2001) (hereafter F01). Here, $R$ is the radius, $\mu$ the actual surface brightness (in mag arcsec$^{-2}$), and $\mu_D$ the same quantity obtained by the extrapolation of $\mu(R)$ in the inner disk, where the exponential decline can be followed unambiguously.

What is $\tau(R)$? Does it depend on the observed wavelength? These questions, of fundamental importance to understanding this phenomenon, motivated the present paper and are only partially answered. Nevertheless, the observations reported here will contribute to achieving a final description, which is difficult to assess because truncations (or ``breaks'') take place at a very low brightness in regions with a very low S/N ratio. The different theoretical models currently proposed are still far from obtain the functional form of $\tau(R)$, but they should be able to do so eventually. Excellent reviews of the question have already been made, such as those by van der Kruit (2001) and Pohlen et al. (2004).

We opted to observe at NIR wavelengths, for the following reasons: firstly, the old stellar population, which makes up the great majority in the galactic disk of spiral galaxies, can be better traced, as it emits the bulk of its flux at NIR. There is also a contribution to the observed flux from the young population and from AGB stars, but this tends to be restricted  spatially to specific areas within the disk (e.g. spiral arms and bulges) and does not change the basic trend in the stellar distribution throughout the disk. Another important argument is that the extinction effects are minimised and so the observed brightness distribution more closely resembles the true stellar distribution, while the stellar flux still largely dominates the measured flux. This is not the case at longer wavelengths, at which the extinction is even lower. Also, the instrumental set-up allows us to reach outer regions on the galactic disks, where truncations are normally found, due to a combination of an excellent dry site, a telescope designed and maintained for NIR observations and cryogenic instrumentation with background contamination controlled via a cold stop.

Most previous studies (for example, Barteldrees \& Dettmar 1994; de Grijs et al. 2001; Kregel et al. 2002; Pohlen et al. 2002a) have been carried out at optical wavelengths. To be complete, we mention the paper by Schwarzkopf \& Dettmar (2000) as it contains a sample of 110 highly inclined galaxies, containing also K$_s$ data for 41 of the galaxies, but most of these are probably not deep enough. 

As a complement to this large amount of statistical data, detailed studies of the truncation curve are required. The observations reported in the present paper and in F01 are considerably deeper than those of any previous study dealing with NIR photometry on the outskirts of spirals, here reaching values of 20-23 mag arcsec$^{-2}$ in J and 18.5-21 mag arcsec$^{-2}$ in K$_s$, which correspond to a face-on surface brightness in the range of 25-26 mag arcsec$^{-2}$ in J and 24-25 mag arcsec$^{-2}$ in K$_s$. The increase of limiting magnitudes from edge-on to face-on views depends on the galaxy, mainly due to its inclination and thickness. These sensitivity limits enable us to extend the structural analysis to the outer areas of the disks, at galactocentric radii of comparable size to those achieved in optical studies.

F01 presented NIR observation dealing with truncations, and the present paper can be considered as a continuation of the former. The telescope and the infrared camera proved to be very efficient for investigating the low surface brightness at the outskirts of disks.

More NIR observations are necessary, especially taking into account that optical and NIR observations may differ substantially and that this difference can be crucial. Recent star formation beyond the truncation (or break) radius would have a larger influence in the blue but very little influence in the NIR range. 

Different studies of truncations have used different techniques and so the results presented are not easily comparable. Some groups observe edge-on galaxies, while others consider face-on galaxies. Comparison is especially difficult when the data used are obtained at different wavelengths. For correct identification of the differences between optical and NIR observations a parallel study  of the same galaxies would be preferable, using the same analysis. This would be the best way to determine, definitively, whether there is a clear difference between the results obtained using the different types of wavelengths. We present here the results of our NIR survey which are noticeably deep and interesting in themselves. Observations at these wavelengths can provide vital clues to our understanding of the truncation phenomenon.

\section{The truncation curve in the literature}

Different descriptions for the truncation curve are classified in the schematic plot shown in Fig. 1. The face-on $\mu(R)$ and the truncation curve $\tau(R)$ are plotted for four basic truncation types. This classification concerns the models considered to interpret the observed profiles. We have adopted this scheme of using the face-on $\mu(r)$ in the truncation curve to homogenise data coming from observations of both, face-on and edge-on galaxies. To render the last ones comparable with the former ones, we must deproject them. Deprojected profiles are intended to represent the true structure of the galaxy.

\begin{figure}
\resizebox{\hsize}{!}{\includegraphics{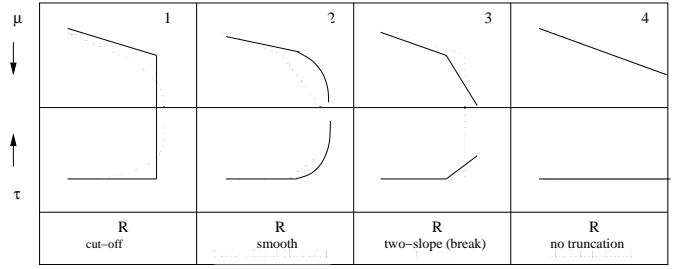}}
\caption{Different ways to describe the truncation phenomenon.}
\label{mur}
\end{figure}

{\it Type 1}. A sharp cut-off is introduced in the model. This idealisation is justified by the fact that the truncation curve interval is relatively narrow.

 The early observations by van der Kruit (1979) and van der Kruit \& Searle (1981a,b) suggested profiles belonging to Type 1 and were interpreted as a sharp cut-off in face-on galaxies that provided a relatively smooth truncation when the galaxy was seen edge-on. A sharp cut-off profile can appear as a smooth decreasing by the line of sight integration which is inherent to the observation of edge-on galaxies. Later on, Wainscoat et al. (1992) and many other works adopted this profile, after the work by van der Kruit and Searle.

{\it Type 2}. A ``smooth'' profile, in which truncation is gradual, i.e. the first derivative is a continuous function and the truncation becomes complete at $R_{tr}$, i.e. $\mu=  \infty$ for $R > R_{tr}$ (within observational errors). There is a transition region of a few kiloparsecs. The edges are rounded in these profiles. 

Note that when we state that truncations are complete, this is an ideal description that is obtained by extrapolation. The zero intensity level is, of course, unreachable. However, in some galaxies the truncation curve is so well defined that ``naked eye'' extrapolation suggests that the truncation is complete.

F01 presented NIR $\mu(R)$ and $\tau(R)$ in agreement with this  ``smooth'' description. The plots in their Fig. 3 for NGC 4013 and NGC 5981 clearly suggest that the first derivative $d\mu/dR$ is a continuous function (in contrast, for example, to type 3). The real curve gradually moves away from the inner exponential to reach very large values. F01 proposed a function of the type
\begin{equation}
  \tau(R) = {{constant} \over {(R_{tr}-R)^n}}
\end{equation}
and after fitting the data, $n$ was found to be close to unity. This function is a very simple one, a power law, for a continuous curve with a continuous derivative function.

The type 2 profile is a smooth version of the classical sharp cut-off type 1 profile, avoiding a discontinuous derivative which is seldom found in nature. Profiles 1 and 2 are essentially the same, with the addition of a taring function, $\tau(R)$, to connect the exponential disk and the sharp cut-off with slope $\infty$ in the type 1 profile. The introduction of a taring function has been considered by Casertano (1983) and by de Grijs et al. (2001), who thus obtained a ``soft cut-off'' profile. However, they adopted other functions that differ from our equation 2.

{\it Type 3}. The radial profiles show a two-slope shape separated by a rather sharp ``break'' radius. At the break radius, the inner exponential is followed by another exponential decline with a different slope. In this case, there is no real truncation, i.e. the truncation is not complete as in the type 2 profiles. This conclusion has also been reached by Pohlen et al. (2002a, 2004) and others.

This profile is very well defined in NGC 5923 (Pohlen et al. 2004). It is noticeable that this galaxy is face-on (in early observations, truncations have been studied in edge-on galaxies) which avoids deprojection effects. The two-slope profile description has also been reported by Erwin et al. (2005) and others.

{\it Type 4}. The galaxy has no break and no truncation. This description corresponds to the early paper by Freeman (1970) where the disk was considered to be purely exponential.

The profiles reported in this paper fit the type 2 ``smooth'' truncation profile, following Eq. 2. Our NIR data are not deep enough to permit a confident exploration, e.g. with sufficient S/N, of the region in which recent optical observations find the second slope after a break.

In this paper, we present the data observed without theoretical preconceptions. It will be followed by a second one showing that the magnetic model matches the statistical properties derived from the data reported in this paper.

\section{Data}

As in F01, the observations were carried out at the 1.5m CST at the Teide Observatory, Tenerife, with the NIR camera CAIN equipped with a $256^2$ NICMOS 3 detector array. The plate scale, with the wide field optics used, was 1.0$\arcsec$/pixel and the effective field of view was 4.3$\arcmin$ x 4.3$\arcmin$. For the CAIN camera, the controller of the detector is based on a San Diego State University architecture, which provides low noise and high stability while permitting fast reading of the whole array, in about 50 milliseconds. In all cases, the read-out mode was the FOWLER mode, in which an equal number of non-destructive readings are taken, without time delays in between, immediately after the initial reset and before the final one, after the selected integration time has elapsed. Thus, the pedestal and final flux levels are better determined, by averaging each group of readings independently. The final signal is then the difference between the last and first averages. The observational characteristics are the same as in F01 (also explained in Castro-Rodr\'{\i}guez \& Garz\'on 2003).
The observations were carried out in 4 campaigns. The weather was generally good, being most of the nights photometric, on which we took the data from the standard stars and galaxies. A consistency check was performed for several bright objects against 2MASS point source catalogue, confirming that our measurements and those of 2MASS are consistent.

The observed galaxies and their main physical parameters are depicted in Table 1. Most of these were obtained from the LEDA database. The position angle (PA) is measured on the image with an automatic routine which delivers high precision values; for most of the galaxies the value obtained coincides with that of the LEDA database, the differences being $\le 2^o$. The orientation of the detector chip with respect to the sky was kept constant throughout the observation runs, so there was no systematic effect due to changes in the instrument configuration. The PA was measured from the North celestial pole, increasing towards the East. Distances were obtained with $H_o=70$ kms$^{-1}$ Mpc$^{-1}$ taking into account the mean heliocentric radial velocity (km/s). All galaxies are edge-on, with an inclination $i= 90^o$ in LEDA. They were also selected to be, as far as possible, free of Milky Way stars in the outer edge and in such a way that their projected size ($D_{25}$) fitted within the FOV. No other physical parameters (other than angular size and galactic latitude) were adopted for the selection; thus, although selection bias cannot be discarded, it is believed improbable.

\begin{table*}
\centering
\caption[ ]{Basic properties of the galaxies.} 
\begin{flushleft}
\begin{tabular}{|l|c|c|c|c|c|c|c|c|} \hline
  Galaxy  &  RA  &  DEC & PA & D & Type &  m$_{abs}$ & log$D_{25}$ & log $v_m$  \\
          & hh mm ss & dd mm ss & d  & Mpc & & & &\\
    (1)   & (2)  &  (3) &(4) &(5) &(6)  &   (7) &  (8) & (9)        \\
\hline
\hline
  NGC 522  & 01 24 45.91 &  09 59 40.5 &  33.3       & 39.0 & Sbc  & -20.53 & 1.44 & 2.252 \\
  NGC 684  & 01 50 14.03 &  27 38 44.2 &  88.8 & 50.4 & Sb   & -21.53 & 1.53 & 2.368 \\
MCG-01-05-047 & 01 52 49.01 & -03 26 51.2 &161.4 & 71.5 & Sc & -21.77 & 1.47& 2.410 \\
  NGC 781  & 02 00 09.02 &  12 39 21.5 &  13.0 & 49.8 & Sab  & -20.87 & 1.18 & \\
  NGC 2654 & 08 49 11.91 &  60 13 13.9 &  65.0 & 19.2 & SBab & -20.09 & 1.62 & 2.295 \\
  UGC 4906 & 09 17 39.94 &  52 59 34.3 &  49.0 & 32.6 & Sa   & -20.26 & 1.30 & 2.228 \\
  NGC 2862 & 09 24 55.10 &  26 46 29.0 & 114.0 & 58.5 & SBbc & -21.44 & 1.41 & 2.464 \\
  NGC 3279 & 10 34 42.61 &  11 11 50.7 & 152.0 & 19.9 & Scd  & -19.27 & 1.44 & 2.208 \\
  NGC 3501 & 11 02 47.35 &  17 59 22.6 &  28.0 & 16.2 & Sc   & -19.05 & 1.54 & 2.147 \\
  NGC 5981 & 15 37 53.55 &  59 23 30.9 & 139.5 & 36.1 & Sc   & -20.61 & 1.43 & 2.424\\
  NGC 6835 & 19 54 33.09 & -12 34 02.5 &  72.0 & 23.0 & SBa  & -19.55 & 1.38 & 1.803\\         
\hline

\end{tabular}

{Columns: (1) galaxy name; (2) and (3) coordinates (2000); (4) position angle; (5) distance, considering $H_o=70 kms^{-1}Mpc^{-1}$; (6) morphological type; (7) absolute B-magnitude; (8) log of apparent diameter (D$_{25}$ in 0.1') and (9) log of maximum velocity rotation (in km/s) from radio observations. All the parameters are obtained from the LEDA database (http://leda.univ-lyon1.fr) except the position angles.}
\end{flushleft}
\end{table*}

At these wavelengths, the night airglow emission is bright and rapidly varying; therefore,  we used the classical observational ON-OFF method with a nodding frequency compatible with that of the background variation in the sky and with an equal integration time on each pointing. The integration time interval was split into several frames to avoid exceeding the linear well depth of the detector due to the high sky flux, in particular in the K$_s$ band. Alternating exposures were taken of the galaxy and the adjacent sky, separated by two minutes at most, for effective sky subtraction. Special care was taken with the flat-fielding and sky subtraction. For the reduction steps, we used several IRAF tasks and the IRAF package REDUCE, developed by R. Peletier.

The observations were carried out in 4 campaigns. Table 2 shows the references for each galaxy: Campaign 1 took place in October 2001, Campaign 2 in March 2002, Campaign 3 between October and November 2002 and Campaign 4 in March 2003. Table 2 also describes the following observational parameters of the galaxies observed: the filter, the individual exposure time per frame (5 or 6 seconds for K$_s$ and H and 30 s for J), the number of frames at each position (4 for J and 20 or 24 for K$_s$ and H), the number of images per galaxy (usually 20 for J and 24 for K$_s$ and H); thus, the total exposure time was 40 (J) or 48 (K$_s$ and H) minutes for the object. The last column shows the detection limit calculated as 3 times the STDDEV over the mean value of the background level, measured directly on the images in areas free from emission of the galaxy object. The typical seeing value was, on average, a little in excess of 1$\arcsec$.  In Fig. 4 we plot contour maps for these galaxies.

\begin{table*}
\centering
\caption[ ]{Observational parameters of the galaxies.} 
\begin{flushleft}
\begin{tabular}{|l|c|c|c|c|c|c|c|} \hline

   Galaxy  &  C &  B  & T$_{exp}$/F &  F/P & Pointing & T & 3$\sigma$ \\
          &  &    & (s)  &   & & ob.(m) & mag/arcsec$^2$\\
    (1)  &  (2) & (3) & (4) & (5) & (6) & (7) & (8) \\
\hline
\hline
  NGC 522  & 3  & J     & 30 &  4 & 20 & 40 &     20.6   \\
           & 3  & K$_s$ &  5 & 24 & 24 & 48 &     21.1   \\
  NGC 684  & 1  & J     & 30 &  4 & 20 & 40 &     20.5   \\
           & 1  & K$_s$ &  5 & 24 & 24 & 48 &     18.6   \\
  MCG-01-05-047 & 3 & J & 30 &  4 & 20 & 40 &     22.7   \\
                & 3 & K$_s$ &  5 & 24 & 24 & 48 & 20.7   \\
  NGC 781  & 3  & J     & 30 &  4 & 20 & 40 &     22.9   \\
           & 3  & K$_s$ &  5 & 24 & 24 & 48 &     20.7   \\
  NGC 2654 & 3  & J     & 30 &  4 & 20 & 40 &     22.6   \\
           & 3  & K$_s$ &  5 & 24 & 24 & 48 &     21.0   \\
  UGC 4906 & 4  & J     & 30 &  4 & 20 & 40 &     22.0   \\
           & 4  & K$_s$ &  5 & 24 & 24 & 48 &     20.9   \\
  NGC 2862 & 3  & J     & 30 &  4 & 20 & 40 &     22.7   \\
           & 3  & K$_s$ &  5 & 24 & 24 & 48 &     20.7   \\
           & 3  & H     &  5 & 24 & 24 & 48 &     21.5   \\
  NGC 3279 & 2  & J     & 30 &  4 & 20 & 40 &     22.0   \\
           & 2  & K$_s$ &  6 & 20 & 24 & 48 &     20.4   \\
  NGC 3501 & 4  & J     & 30 &  4 & 20 & 40 &     22.3   \\
           & 2  & K$_s$ &  6 & 20 & 24 & 48 &     20.4   \\
  NGC 5981 & 2  & J     & 30 &  4 & 20 & 40 &     21.9   \\
           & 2  & K$_s$ &  6 & 20 & 24 & 48 &     20.2   \\
  NGC 6835 & 1  & J     & 30 &  4 & 20 & 40 &     20.1   \\
           & 1  & K$_s$ &  5 & 24 & 24 & 48 &     18.7   \\
\hline
\end{tabular}

Columns: (1) galaxy name; (2) campaign (1, oct. 2001; 2, march 2002; 3, oct.-nov. 2002; 4, march 2003); (3) band; (4) individual exposure time per frame; (5) number of frames at each position; (6) images per galaxy; (7) exposure time for the galaxy and (8) the 3$\sigma$ level of the background variation over its mean value.
\end{flushleft}
\end{table*}

\section{Deprojection}

It is clear that the deprojection technique is critical in the reduction process. In some cases, an initial inspection of the original profiles may suggest a truncation which disappears after deprojection. To assure that the deprojection method works satisfactorily, we carried out two tests:

a) Deprojection was performed by two different, independent methods. The first of these is described in detail in F01, while the second is based on Abel's integral equation (Binney \& Tremaine 1987). Despite the fact that the numerical calculations involved in these methods are intrinsically different, the results obtained by the F01 method and by using Abel's integration are almost indistinguishable. The F01 method was subsequently adopted, because of the few assumptions needed, and to use the same reduction process for the sake of coherence when merging the present data with those obtained in F01. Thus, a larger and more homogeneous statistical sample is ensured. Moreover, the F01 method can be generalised to more realistic conditions, including extinction.

b) We have tested that the projection of the deprojected profile (x points) perfectly matched the original profile ($\triangle$ points) in a zero order test (Fig. 2). A deprojected profile corresponded to only one original profile and an original profile corresponded to only one deprojected profile. For example, our smooth deprojected profiles were not produced by a face-on profile with a sharp cut-off. 

\begin{figure}
\resizebox{\hsize}{!}{\includegraphics{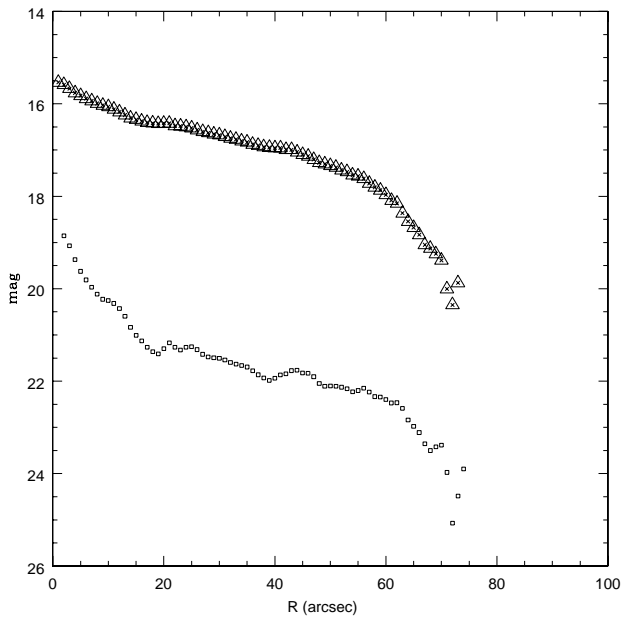}}
\caption{Test for the deprojection method. Symbols are: $\triangle$, input curve; $\square$, deprojected curve; $x$, projection of the deprojected curve.}
\label{pru}
\end{figure}

We took the z-integrated profiles as the original input profile for the deprojection method, i.e. those profiles obtained by summing all the input radial profiles for each z. These z-integrated profiles were obtained as follows: first the galaxies were rotated to make the major axis perpendicular to the ordinate axis. From I(r,z), where I is surface brightness, r is the abscissa and z the ordinate, the z-integrated profile is then obtained as $I(r) = \Sigma_{z=a}^b I(r,z)$, where 
$a$ and $b$ are the limits of a box which contains the $3\sigma$ 
isophote of galaxies. We have not masked any foreground stars. We thus create an artificial profile. If we wished to obtain the truncation curve as a function of z, this z-integration would prevent it. This could be interesting, for instance, to see if the truncation were different for the thin and the thick disks, which is especially important in early type spirals. Moreover, in early type galaxies the extended bulge could smear out a possible truncation in a z-integrated profile. This cannot be undertaken when using face-on observations.

Certainly, by z-integrating we cannot obtain the emission per unit volume in the galaxy. However, this procedure has two important advantages: firstly, we can work with a profile which has a much better signal/noise ratio, and secondly, the effect of the large extinction dependence on z is greatly reduced. Suppose, for instance, a dust lane for z=0 or around z=0. The row z=0 would give meaningless results because of severe extinction. By summing all the z rows, the effects of localized dust lanes are reduced.

The z-integration apparently leads to a loss of information. However, this effect is irrelevant if the final product of the deprojection method is the face-on view. In Fig. 3 we plot both the z-integrated and the deprojected profiles for a selected galaxy, NGC 2862, for the three filters to illustrate the difference.

\begin{figure}
\resizebox{\hsize}{!}{\includegraphics{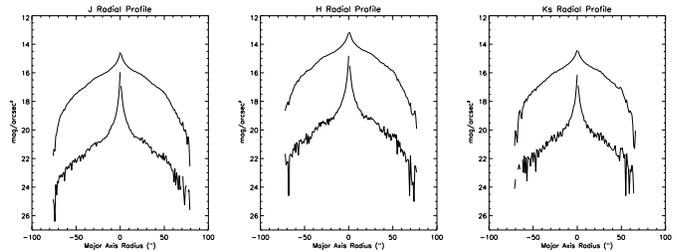}}
\caption{NGC 2862: z-integrated profile (upper) and deprojected profile 
(lower) for each filter.}
\label{perfil}
\end{figure}

It might be thought that in order to obtain the face-on view the sequence ``deprojection, then z-integration'' is not equivalent to the sequence ``z-integrating, then deprojection''. From the description of the deprojection method in F01 it can be demonstrated that the two sequences are equivalent. This is shown in Appendix A.

Therefore, the method itself does not introduce important errors (as concluded in F01). However, important errors could be introduced if the required assumptions are not accomplished. These assumptions are axisymmetry, the absence of extinction, and an inclination equal to 90$^o$. In other words, the deprojected profile need not necessarily coincide with the real face-on profile. They only coincide when the required assumptions are met.

With respect to axisymmetry, it is clear that when the galaxy lacks this property then deprojection cannot reproduce the real face-on view. This source of potential errors is inevitable in this and in any other deprojection method. The problem has infinite solutions, as from a 2D distribution it is not possible to obtain a 3D one without certain assumptions. Fortunately, in the galaxies observed, the noticeable east-west symmetry suggests that they are reasonably axisymmetric.

An estimation of this error is not easy. NIR data of face-on galaxies have not been reported in the literature up to these large radii, so we cannot use additional real data for this estimate. The only galaxy for which we have at our disposal NIR data at large radii is the Milky Way. Wainscoat et al. (1992) have modelled the Milky Way in order to reproduce IRAS data, including NIR arms which are the mean sources of non-axisymmetries. The over-density due to one NIR arm is very small. From these data about the relative importance of arm with respect the unperturbed disc and taking into account the contribution to the flux of each spectral type, from Porcel (1997) we obtain that the contribution is mainly due to stars of spectral types B8 V to F5 V and is only 1\%. When the propagation of errors in the deprojection method is considered, this figure becomes representative of the error introduced by one-arm asymmetries by the deprojection method. Galaxies reported here could be less symmetric than the Milky Way, but we estimate that this kind of error can be ignored.

Extinction is another inevitable source of errors, particularly if there is a non-axisymmetric distribution of dust. However, the problems introduced by extinction are reduced in the present approach because we work in the NIR and because prior z-integration avoids the effects of inhomogeneities on the distribution of dust, e.g. dust lanes in the plane. 

We used only galaxies with i=90$^o$, following LEDA. However, the LEDA galaxies may deviate typically 2$^o$ from edge-on, which could be a source of errors. Our galaxies have a minor-axis of $70\arcsec \sin{2^o}= 2.4\arcsec$, which is higher than our resolution about $1\arcsec$. In Appendix B we show that inclinations $i > 88^o$ do not introduce noticeable errors.

Another important step in the analysis was to calculate the slope of the exponential function of the form $I(R) = I_0 \exp{(-R/h)}$ in the inner part. This parameter $h$ is present in the definition of $\tau(R)$ and it is necessary to be able to estimate $R_{tr}/h$, another parameter that is derived in all the studies made of truncations/breaks. The region for calculating $h$ was decided upon by visual estimation. h is calculated in the deprojected profiles. Noise and intrinsic variations in surface brightness do not permit the determination of $h$ with an error of less than about 10$\%$.

The use of a visual decision regarding the interval for calculating $h$ does not introduce large uncertainties, because of the broad region in which the exponential behaviour provides a good fit. If the photometric profiles are divided into bulge, exponential and truncation shapes (with no sharp transitions) the choice of an excessively large interval would provide an underestimation of $h$, while the choice of one that was excessively small would give large errors. But excluding these two extreme cases, the value of $h$ is nearly independent of the choice of the interval. Therefore, this subjective choice does not introduce additional errors.

The main results are presented in Table 3, for each side of the galaxies. After the deprojection process, by fitting the exponential profile, we obtain $\mu_o$, the central surface brightness of the disk, and $h$, the radial scale length. Then we estimate $R_{tr}$, the truncation radius, and $n$, the exponent in equation (2). We also calculate $R_{tr}/h$.

\begin{table*}
\centering
\caption[ ]{\vspace{-0.2cm}Photometrical and structural parameters of the target list}
\begin{flushleft}
\begin{tabular}{|l|c|c|c|c|c|c|c|c|c|} \hline
  Galaxy & Passband & Side & $\mu_o$ & $h $ & $ h $ & $R_{tr}$ & $R_{tr} $ & n & $R_{tr}/h$ \\
         &          &      &(mag/arcsec$^2$)& (kpc)& (arcsec) & (kpc) & (arcsec) & &  \\
(1) & (2) & (3) & (4) & (5) & (6) & (7) & (8) & (9) & (10) \\
\hline
\hline
%%% A partir de aqui van los datos de los objetos:
  NGC 522       & J     & SW/NE & 20.6/20.6 &  7.2/6.1  & 38/32 & 14/14 & 73/72  & 1.0/1.0 & 1.9/2.2 \\
                & K$_s$ & SW/NE & 21.1/20.9 &  8.0/6.3  & 42/33 & 11/11 & 59/60  & 0.5/0.7 & 1.4/1.8 \\
  NGC 684       & J     & NW/SE & 18.8/18.6 &  5.3/4.6  & 22/19 &       &        &         &         \\
                & K$_s$ & NW/SE & 16.8/17.3 &  3.7/4.1  & 15/17 &       &        &         &         \\
  MCG-01-05-047 & J     & NW/SE & 20.9/20.9 & 10.4/11.3 & 30/33 & 30/29 & 85/84  & 0.6/0.7 & 2.8/2.6 \\
                & K$_s$ & NW/SE & 19.5/19.6 &  8.4/9.4  & 24/27 & 28/27 & 82/79  & 1.0/1.0 & 3.4/2.9 \\
  NGC 781       & J     & SW/NE & 19.4/19.8 &  2.6/3.1  & 11/13 & 11/11 & 46/45  & 0.9/0.8 & 4.2/3.5 \\
                & K$_s$ & SW/NE & 18.3/18.0 &  2.6/2.3  & 11/9  & 10/10 & 41/40  & 0.5/0.3 & 3.5/4.3 \\
  NGC 2654      & J     & NE    & 18.9      &  1.5      & 17    &       &        &         &         \\
                & K$_s$ & NE    & 17.3      &  1.4      & 15    &       &        &         &         \\
  UGC 4906      & J     & SW/NE & 18.3/18.3 &  1.9/1.7  & 12/11 &       &        &         &         \\
                & K$_s$ & SW/NE & 17.3/17.3 &  1.7/1.7  & 11/10 &       &        &         &         \\
  NGC 2862      & J     & NW/SE & 19.5/19.6 &  5.9/6.0  & 21/21 & 22/22 & 77/76  & 1.1/0.9 & 3.7/3.6 \\
                & K$_s$ & NW/SE & 18.5/18.5 &  5.9/5.6  & 21/20 & 19/18 & 66/65  & 0.8/0.7 & 3.2/3.3 \\
                & H     & NW/SE & 18.7/18.7 &  6.4/5.8  & 23/20 & 21/20 & 74/69  & 1.1/0.3 & 3.3/3.4 \\
  NGC 3279      & J     & NW/SE & 19.9/19.6 &  4.7/3.1  & 48/32 &  8/8  & 79/82  & 1.5/1.0 & 1.6/2.6 \\
                & K$_s$ & NW/SE & 18.5/18.3 &  3.8/2.9  & 39/30 &  7/7  & 69/74  & 1.2/1.2 & 1.8/2.4 \\
  NGC 3501      & J     & SW/NE & 20.2/19.9 &  2.1/1.7  & 27/22 &  8/9  &100/110 &         & 3.7/5.1 \\
                & K$_s$ & SW/NE & 18.1/18.0 &  1.6/1.6  & 21/20 &  6/   & 82/    & 0.5/    & 3.9/    \\
  NGC 5981      & J     & NW/SE & 19.4/19.3 &  4.3/4.2  & 25/24 & 15/14 & 88/79  & 0.9/0.9 & 3.6/3.3 \\
                & K$_s$ & NW/SE & 18.1/18.0 &  4.0/3.7  & 23/21 & 13/12 & 74/70  & 0.9/0.0 & 3.2/3.3 \\
  NGC 6835      & J     & NW/SE & 18.1/18.1 &  1.8/1.8  & 16/16 &       &        &         &         \\
                & K$_s$ & NW/SE & 17.4/17.4 &  1.8/1.8  & 16/16 &       &        &         &         \\

\hline
\end{tabular}

Columns: (1) galaxy name; (2) filter; (3) side of the galaxy; (4) central surface brightness; (5) \& (6) radial scale-length in kpc and in arcsec; (7) \& (8) truncation radius in kpc and arcsec; (9) parameter $n$ obtaining from Eq. (2); (10) ratio between truncation radius and scale-length. All values are deprojected (face-on).
\end{flushleft}
\end{table*}

\section{Results}

The galaxies observed can be grouped taking into account the existence of truncation,  as follows:

a) Truncated galaxies: NGC 522, MCG-01-05-047, NGC 781, NGC 2862, NGC 3279, NGC 3501 and NGC 5981. Our study is mainly focused on this group.

b) Untruncated galaxies: NGC 684, UGC 4906, NGC 2654 and NGC 6835.
 Of course we cannot assure that these are not truncated, but only state that no truncation was observed for face-on values of J$<$23 mag/arcsec$^2$ in NGC 684, for J$<$24 mag/arcsec$^2$ in UGC 4906, for J$<$25 mag/arcsec$^2$ in NGC 2654 or for J$<$22 mag/arcsec$^2$ in NGC 6835. We refer to the J filter limits since in these galaxies the J band data covers a more extended region in the object, as can be seen in Fig. 4. In consequence we did not detect truncations out to 3.5 radial scale-lengths for NGC 684, 5.0 for UGC 4906, 5.9 for NGC 2654 or 3.3 for NGC 6835.

We deduce that approximately 2/3 of the considered galaxies are truncated, a sufficiently large value to highlight, once more, the importance of this effect. Although this figure must be approached with some caution, the overall conclusion must be that truncations/breaks are a fairly common feature of normal spirals. It is also evident that we cannot speak of untruncated galaxies, but merely of galaxies for which truncation is not observed within the limits of our observational capabilities.

In Fig. 5, we plot the deprojected profiles for these galaxies. Fig. 5 also show $\tau(R)$ obtained from Eq. 1 for these galaxies and for two galaxies (MCG-01-05-047 and NGC 5981), included as examples, we show the fitted truncation curve.

Table 3 shows that the truncation radii obtained with the J filter is always larger than the value obtained with the K$_s$ filter, the ratio being $R_{tr}(J)/R_{tr}(K_s) = 1.15$. The dispersion of the data is 0.04, but our error in determining this quantity is larger, at about 0.5. 

The identification performed is not free of definition problems. Let us assume a galaxy with an extended (not sharp) ``break'' region and with the second outer exponential lying beyond observation capabilities. This could have been classified as ``smooth'' despite being a ``break'' galaxy.  The data for some of our galaxies strongly suggest that the ``smooth'' description is physically different from the two-slope type, but 
this identification is not easy. Subjective inspection is probably essential to determine whether the change in the slope is gradual or sharp or, equivalently, whether the first derivative is a continuous or a discontinuous function. Given the mounting body of evidence provided by Pohlen et al. (2002a, 2004), P\'erez (2004), Erwin et al. (2005), Trujillo \& Pohlen (2005) and others in obtaining clear two-slope profiles, we could have fit the data to this model. But considering that our data do not permit the exploration of the second-slope region and considering that the NIR and the optical profiles do not necessarily coincide, we follow an independent approach.

In the previous paper by F01, the deprojected profiles of NGC 4013 (both J and $K_s$, both sides), NGC 4217 ($K_s$, left side) and NGC 5981 ($K_s$, right side) are clear examples of a gradual decline and a continuous first derivative. Notice that NGC 5981 is present in both papers (in the first one only in K$_s$). The K$_s$-images are very similar as expected, but in the first campaign better observing conditions allowed us to reach slightly more distant radii. The difficulties inherent in subjective discrimination cannot be easily surmounted by statistical methods, as the break point should be determined manually. Thus, we prefer the  ``smooth'' truncation, type-2 profiles with gradual decline and gradual variation of the first derivative.

We then fitted the observation points to the function in Eq. 2. $R_{tr}$ was obtained by extrapolation and the ``constant'' and the value of $n$ by standard fitting methods. The $n$ value is particularly interesting. The results are shown in Table 3.

This fitting features the compromise of a gradual transition from a pure exponential to a continuously bending profile. This is, however, provided automatically by the  function given in Eq. (2). $\tau(R)$ becomes negligible for values of $R$ not far from $R_{tr}$. For example, assuming n=1 as a typical value, it is found that $\tau$ is only 10\% of the value for $R= 0.95 R_{tr}$, for a value of $R$ one radial length less than $R_{tr}$. Therefore, the result is practically independent of the starting point in the fitting (provided the bulge zone is excluded). The real curves in Fig. 5 show how fast $\tau$ decreases far from the truncation radius.

The mean value of $n$ is 0.8 with $\sigma =0.3$. This supports the conclusion by F01 who also found a coefficient close to unity. Therefore, we propose
\begin{equation}
  \tau(R) \propto {1 \over {(R_{tr}-R)^ {0.8}}}.
\end{equation}

Notice that the face-on and edge-on truncation curves differ but $R_{tr}$ is the same (Fig. 3). Therefore, this value is unaffected by the use of a deprojection technique. The $R_{tr}$ values obtained in other studies without deprojection are equally valid for statistical studies.

The $R_{tr}$ and $n$ parameters are obtained independently. The value of $R_{tr}$ tells us ``where'' the truncation takes place and the values of $n$ tell us ``how sharp'' the truncation is.

\section{Conclusions}

In order to obtain more information about the truncation of the stellar disk, it is necessary to incorporate observations in the NIR, which is the best tracer of the old population and relatively free of extinction problems, and to use larger telescopes. On the other hand, observations in the optical and in the NIR should be considered complementary, as both wavelength ranges may give information about different aspects of the feature. In fact, they may differ not only in the value of certain fitting parameters, but in the functional form of the truncation curve itself. For example, beyond the truncation (or break) radius, non negligible star formation could affect the young population distribution but not that of the old population. If stars did not move from their birthplace, a difference in the blue and the NIR profiles would not be expected, but stars can move.

If there is a (smooth and complete) truncation in the IR, as shown by F01 and this study, and a sharp break in the optical, as shown by Pohlen et al. (2004), this is not necessarily a contradiction; it could be the sign of  differing behaviour patterns in the optical and in the near infrared, for dynamic reasons. We choose the traditional description introduced by van der Kruit (1987), introducing a taring function to smooth out the sharp cut-off, also called a soft cut-off, and propose a truncation curve proportional to $(R_{tr} - R)^{-n}$, with $n= 0.8 \pm 0.3$, a value close to 1, i.e. it becomes complete at $R=R_{tr}$. This is the profile of type 2, labelled ``smooth'' in Fig. 1.

The type 3 profile is clearly different from that of type 2, either due to the different information provided by the NIR and the optical, or due to insufficient coverage of the NIR observations. In the light of the former consideration, we have chosen the simplest fit suggested by our data, but they are not deep enough to conclude that an external region with a second slope does not exists.

Unfortunately, there are not many galaxies measured in both the optical and the NIR. One of them is NGC 522, which was measured in the optical by Pohlen et al. (2002b). This galaxy is a very good example of a two-slope with a narrow break region in between, when measured in the optical. However, as can be seen in Fig. 5, the J-profile on the right hand side seems to be smooth. The left side as well as the right Ks side could be described as a two-slope. The data does not allow us to clearly distinguish between these two types. Hence, no clear conclusions can be obtained from this comparison. This important point should be addressed in a future study.

As the radial scale height is a natural unit of length in a spiral galaxy, $R_{tr}/h$ is a parameter characterising truncations. For these galaxies, we found the following values for  $R_{tr}/h$: 3.2 for J and 2.9 for K$_s$, with $\sigma=0.9$ and 0.8 respectively for each filter. The colour dependences of $R_{tr}$ will be discussed in Paper II. These values are in reasonable agreement with previous estimations. For  comparison, it should be taken into account that the difference between the radius at which the truncation curve leaves the inner exponential and the radius at which the truncation finishes is of the order of 20-30$\arcsec$, and that in the break description only the first value is meaningful.

It is not easy to establish how often galaxies are truncated, due to the difficulty in determining whether a given galaxy is non truncated or whether $R_{tr}$ is too distant to be observed. If we consider 7 out of 11 galaxies truncated, this is equivalent to approximately 2/3. However, this figure should be considered with some caution as our sample is small. However, we confirm that the frequency of truncated galaxies is high, which makes this topic of crucial importance in our understanding of the evolution of disk galaxies.

\begin{acknowledgements}
This paper has been supported by the ``Plan Andaluz de Investigaci\'on'' (FQM-108) and by the ``Secretar\'{\i}a de Estado de Pol\'{\i}tica Cient\'{\i}fica y Tecnol\'ogica'' (AYA2004-08251-C02-02, ESP2004-06870-C02-02). This research has made use of the Hyper Leda database.
\end{acknowledgements}

\newpage

\begin{figure*}
\centering
\includegraphics[width=11.5cm]{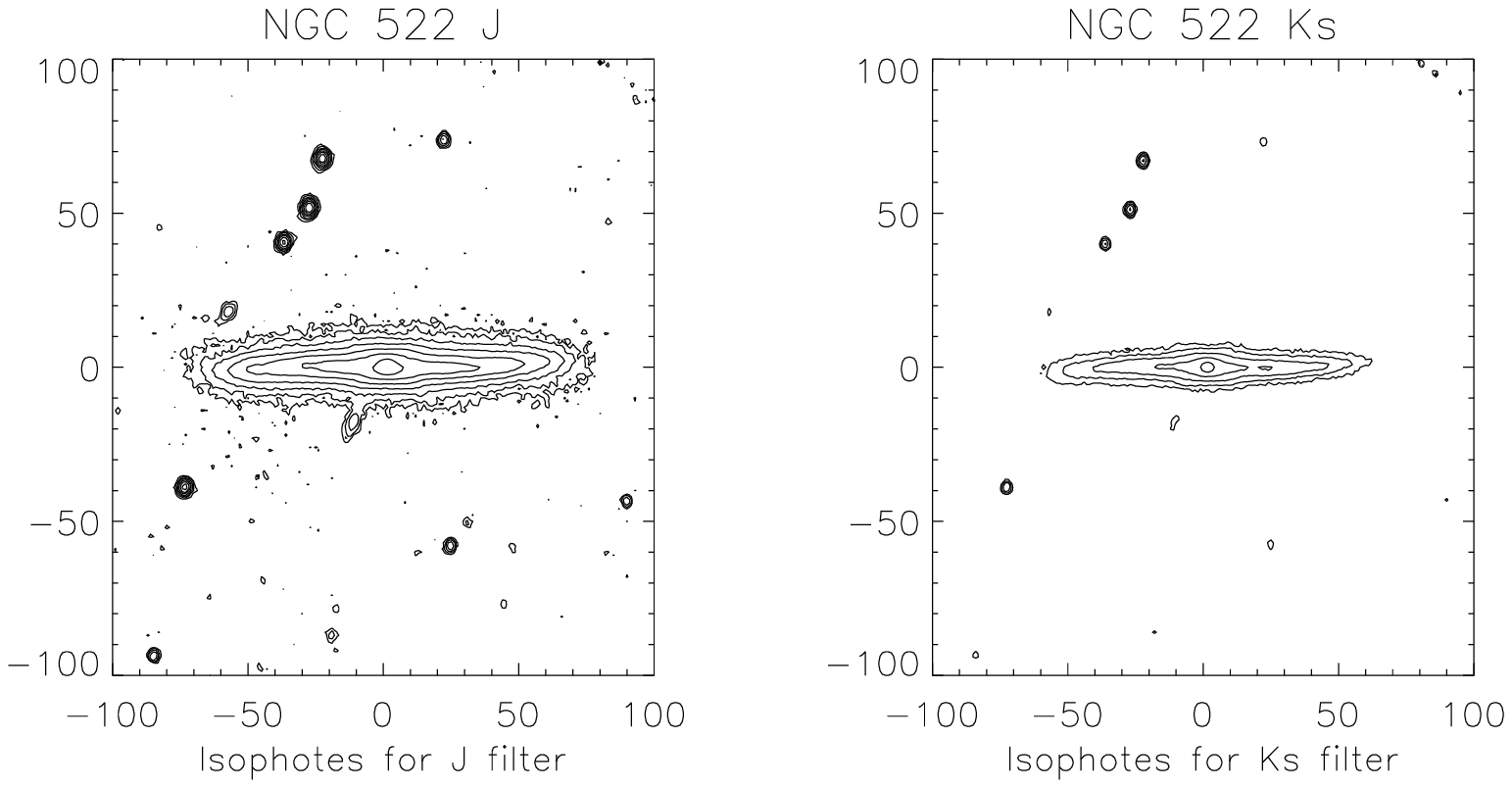}
\includegraphics[width=11.5cm]{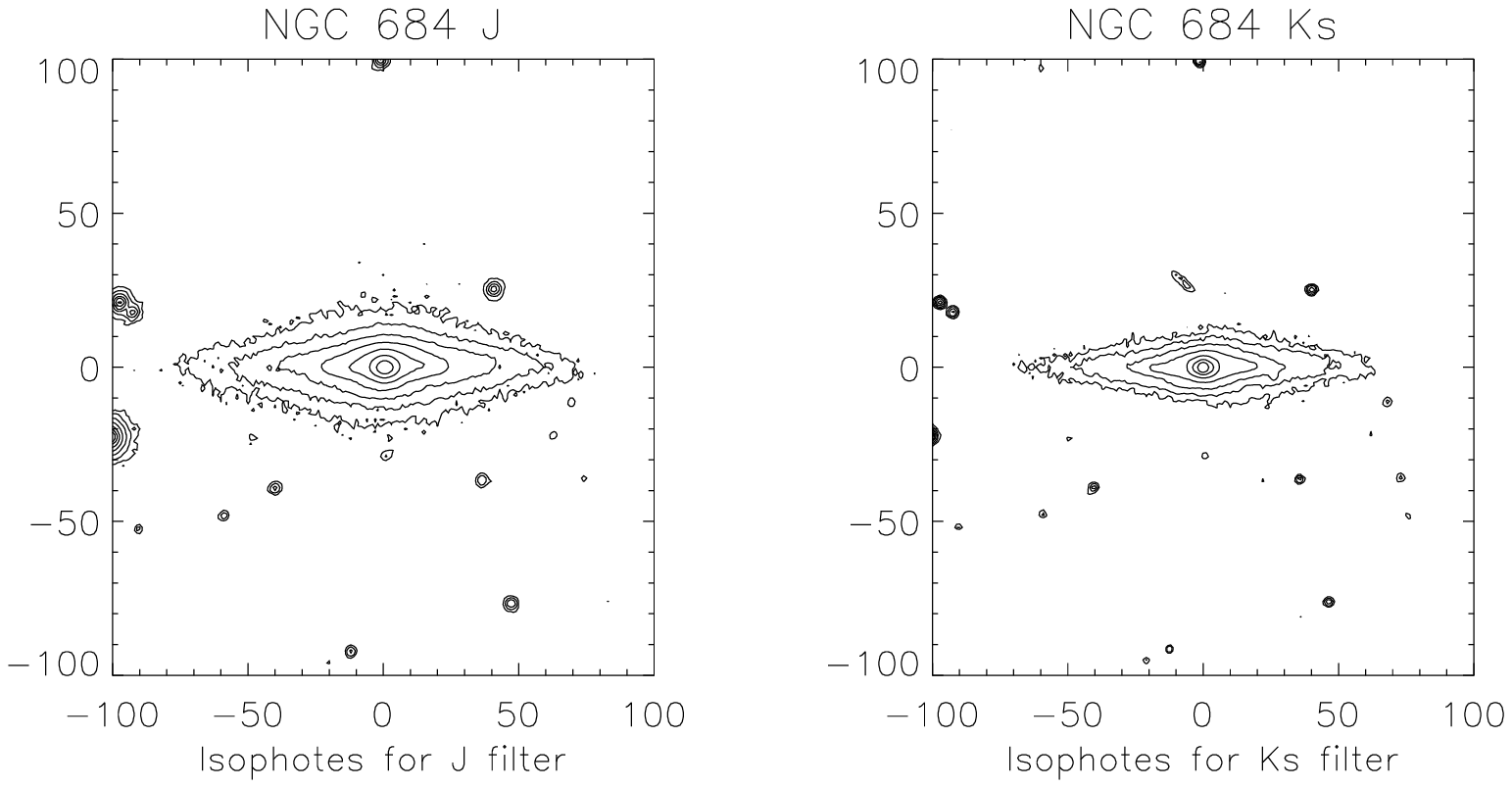}
\includegraphics[width=11.5cm]{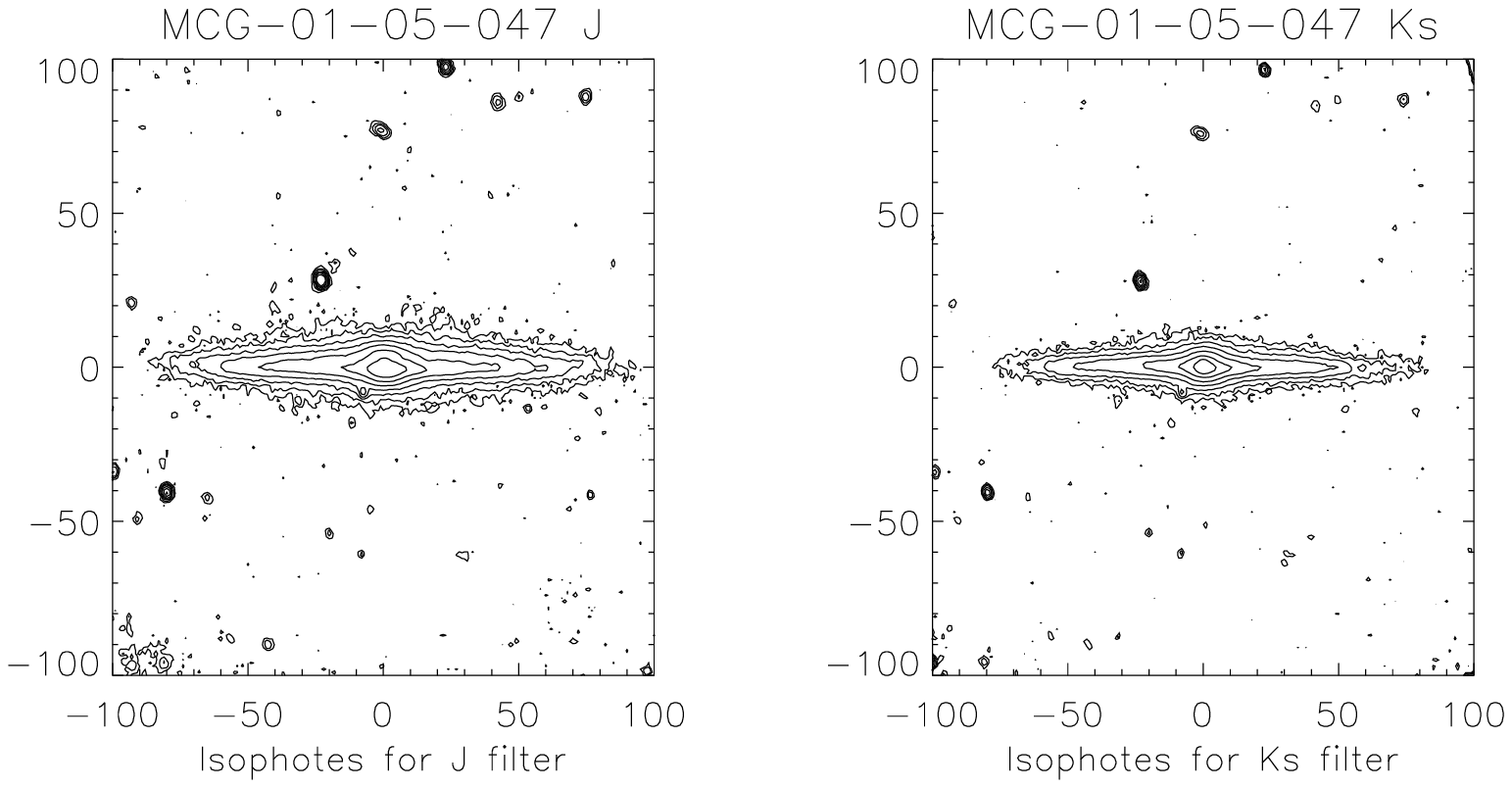}
\includegraphics[width=11.5cm]{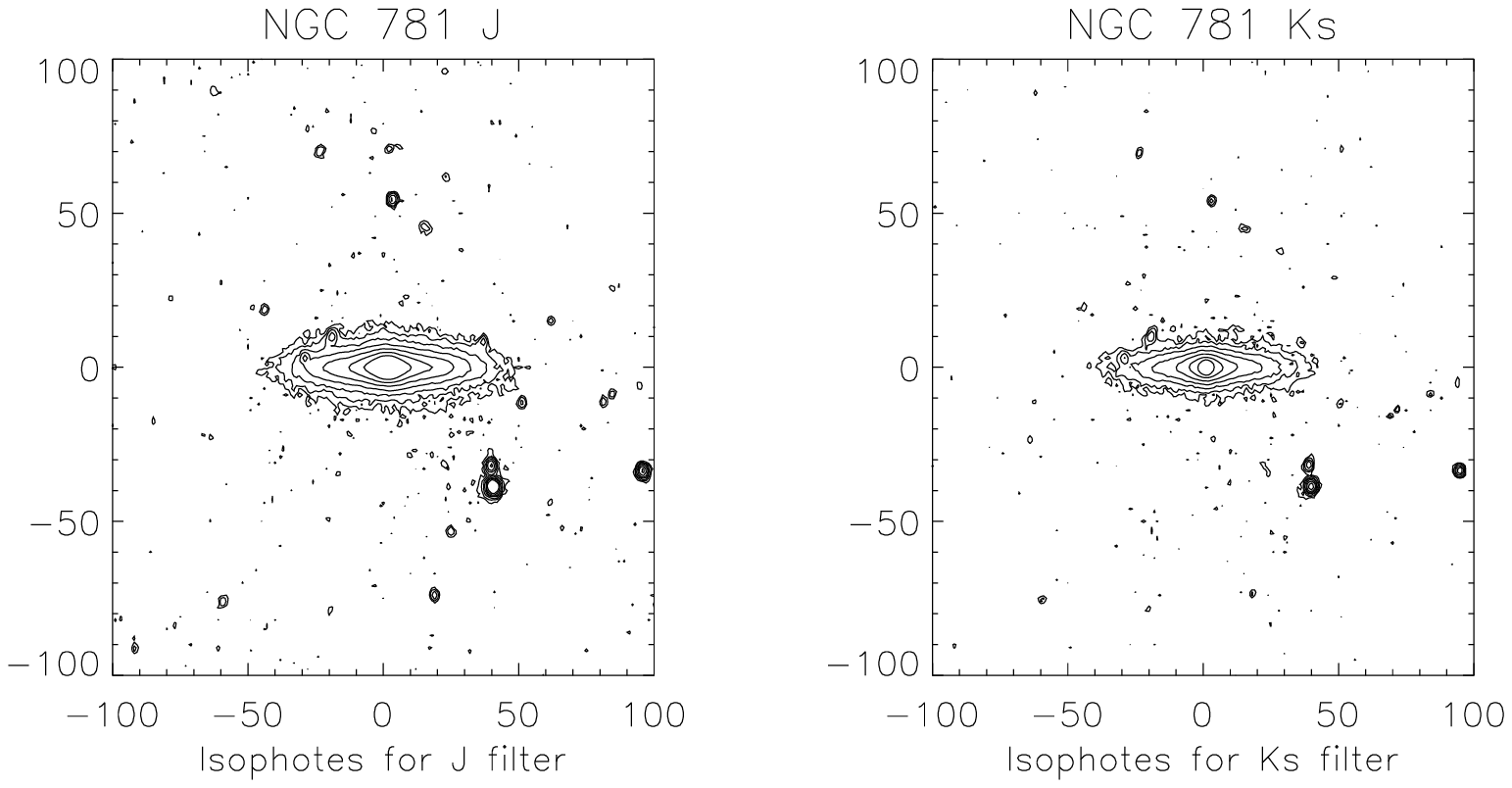}
\caption{Contour maps for the observed galaxies for J, H and Ks 
filters. The isophotes are equidistant (in units of n*3$\sigma$ equiv. 
to a step of +0.75 mag/arcsec$^2$) starting  at our noise level (see 
Table 2). Galaxies are rotated, but N is closer to the top and E to 
the left.}
\label{depro}
\end{figure*}

\newpage

\begin{figure*}
\centering
\includegraphics[width=17cm]{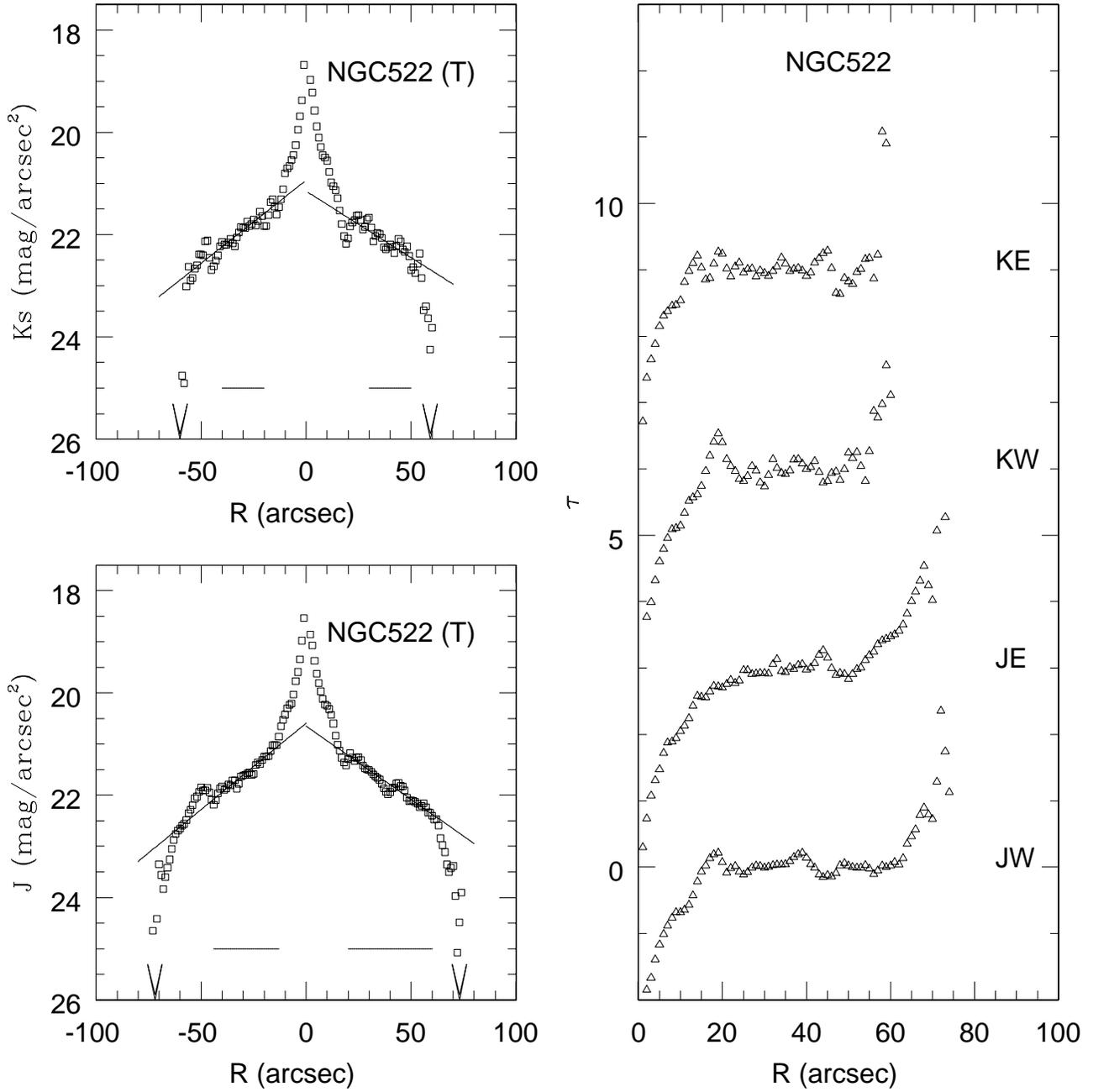}
\caption{Deprojected profiles for all galaxies. Left: The solid line is the exponential fit with the parameters in Table 3. Horizontal bars indicate the region in which the radial scale length was calculated and the arrow the truncation radius.(T) is for truncated galaxies and (U) for untruncated ones. Right: For each of these galaxies we have represented the truncation curve ($\tau$) from the equation 1 and, as an example, we have plot the fitted function of Eq. 2 for two of them.}
\label{truncadas}
\end{figure*}

\begin{figure*}
\setcounter{figure}{3}
\centering
\includegraphics[width=12cm]{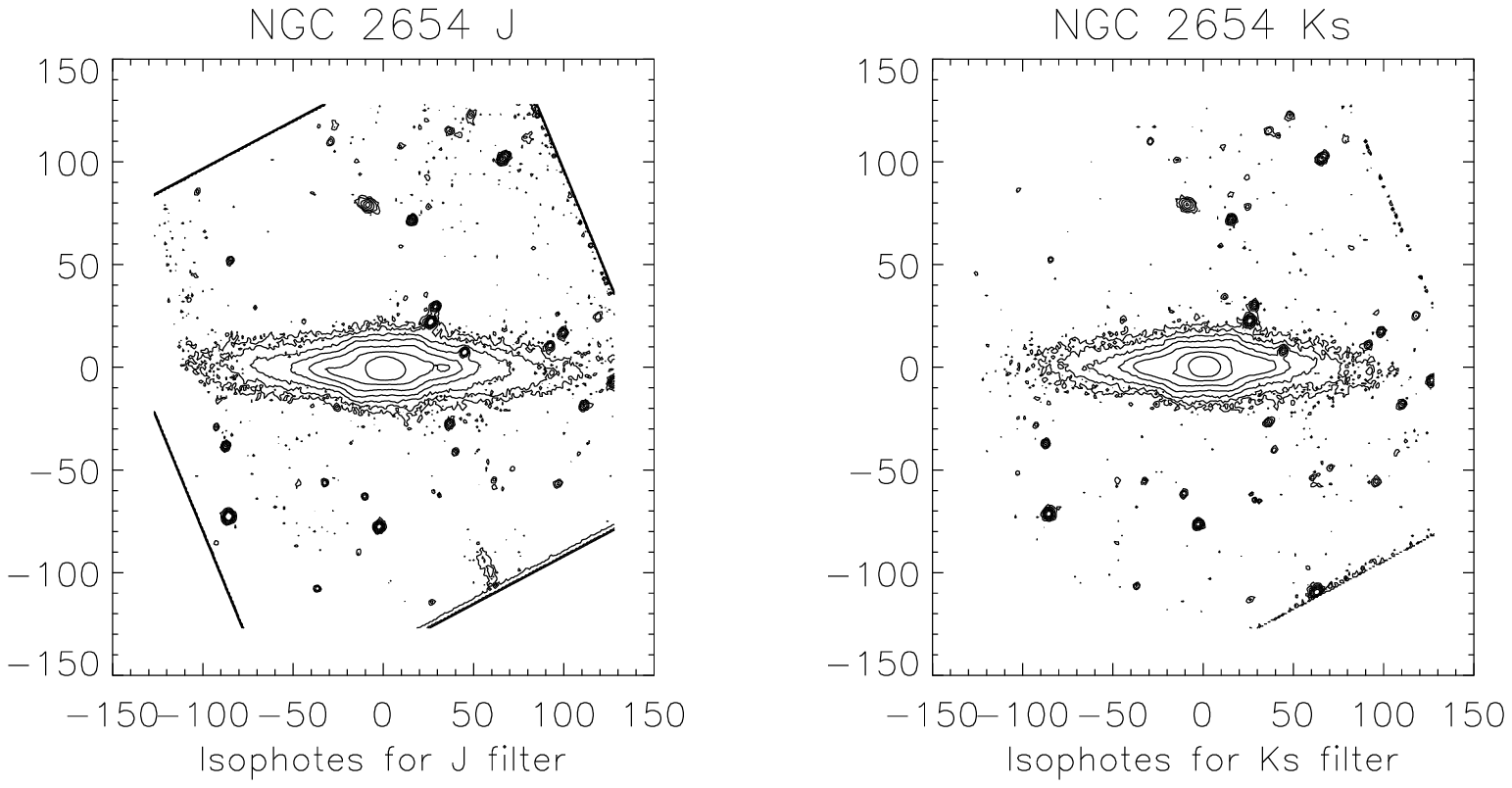}
\includegraphics[width=12cm]{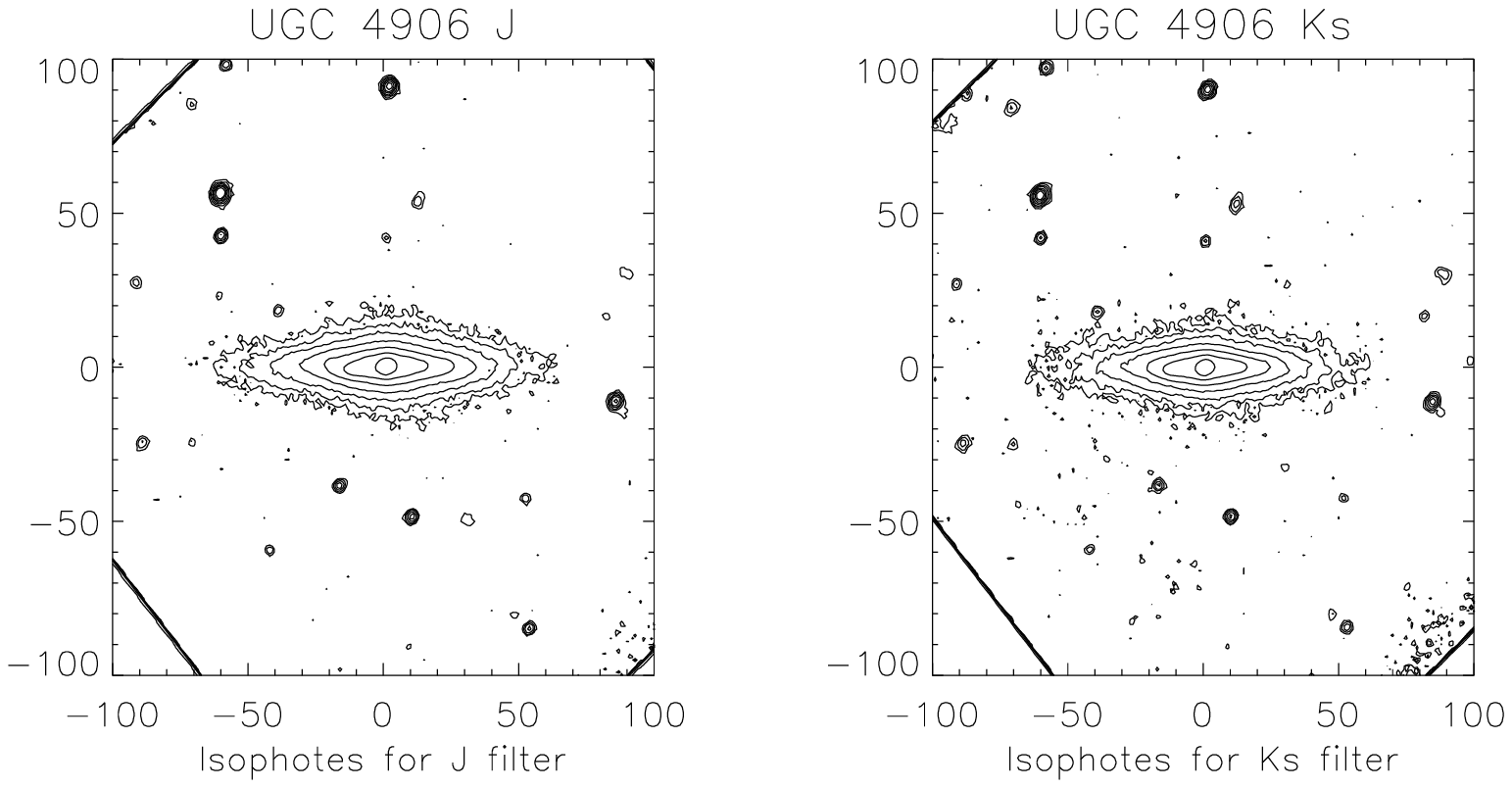}
\includegraphics[width=12cm]{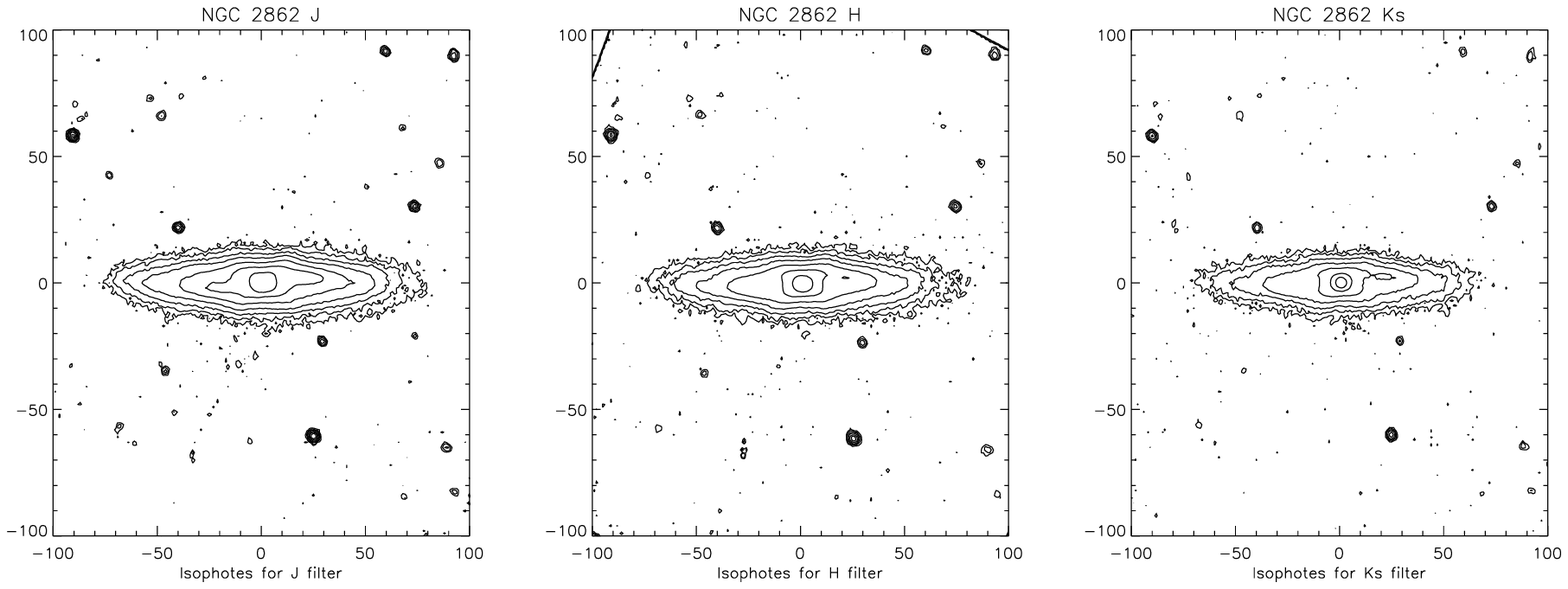}
\includegraphics[width=12cm]{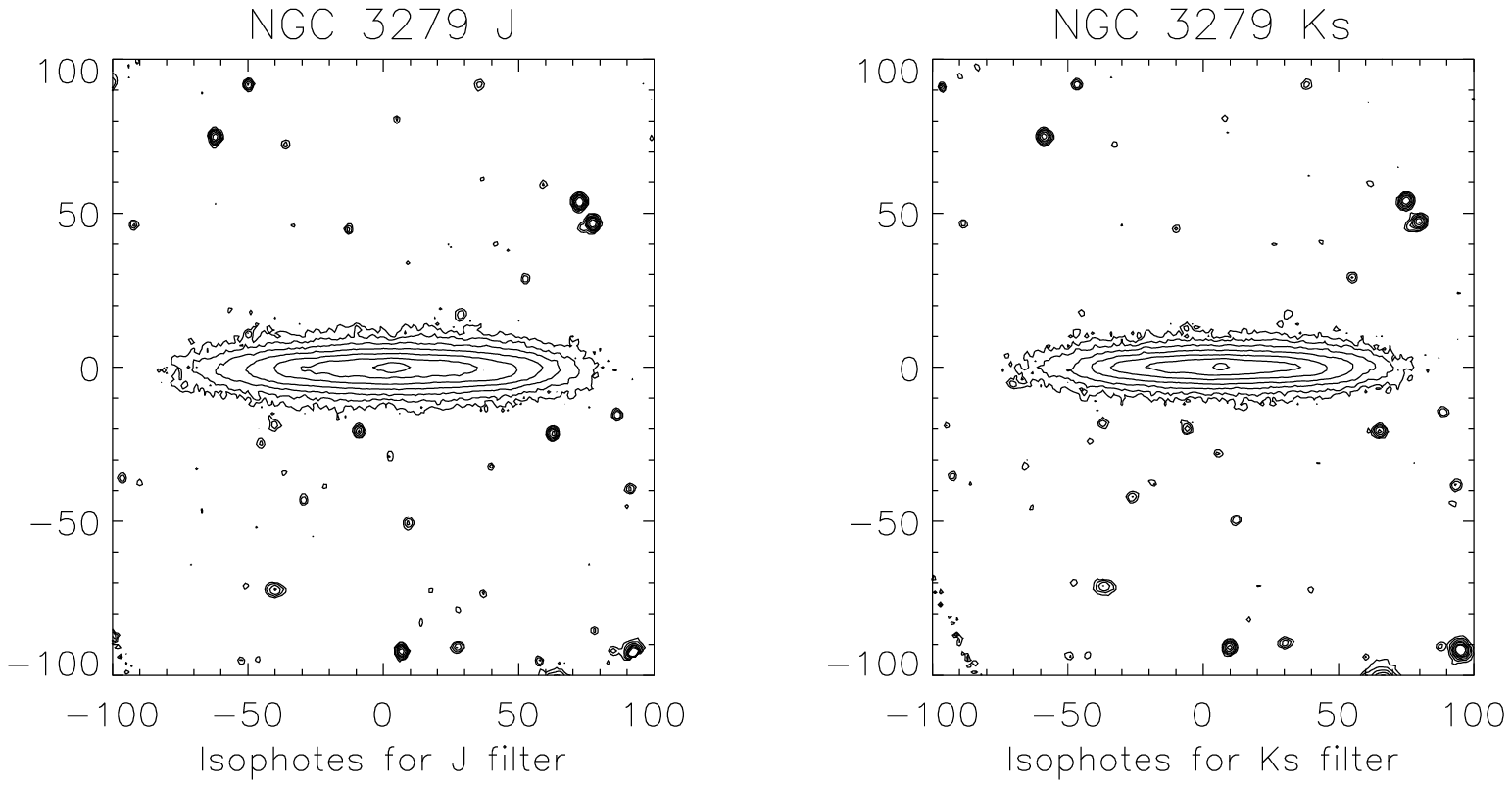}
\caption{continued}
\end{figure*}
\begin{figure*}
\setcounter{figure}{3}
\centering
\includegraphics[width=12cm]{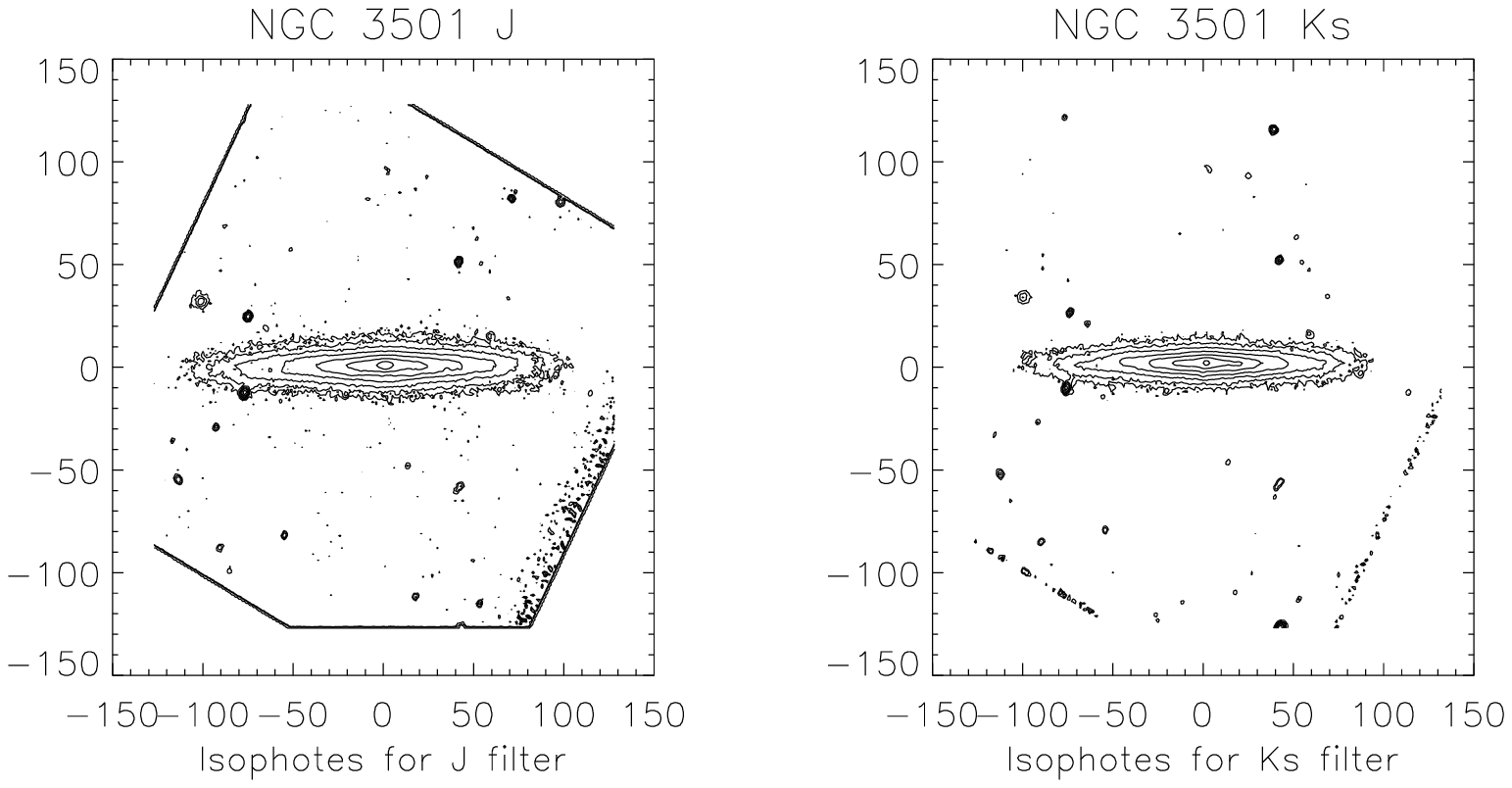}
\includegraphics[width=12cm]{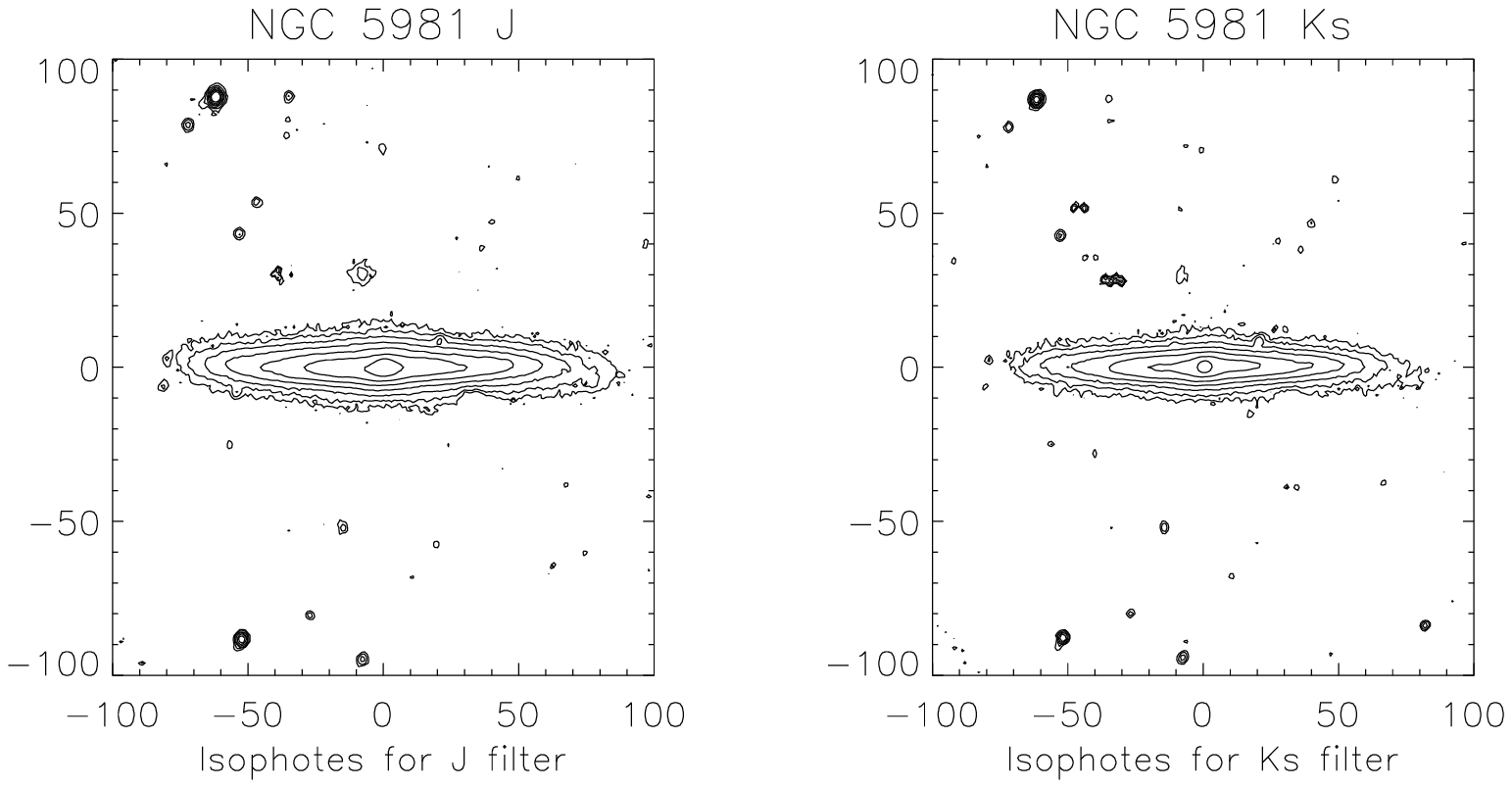}
\includegraphics[width=12cm]{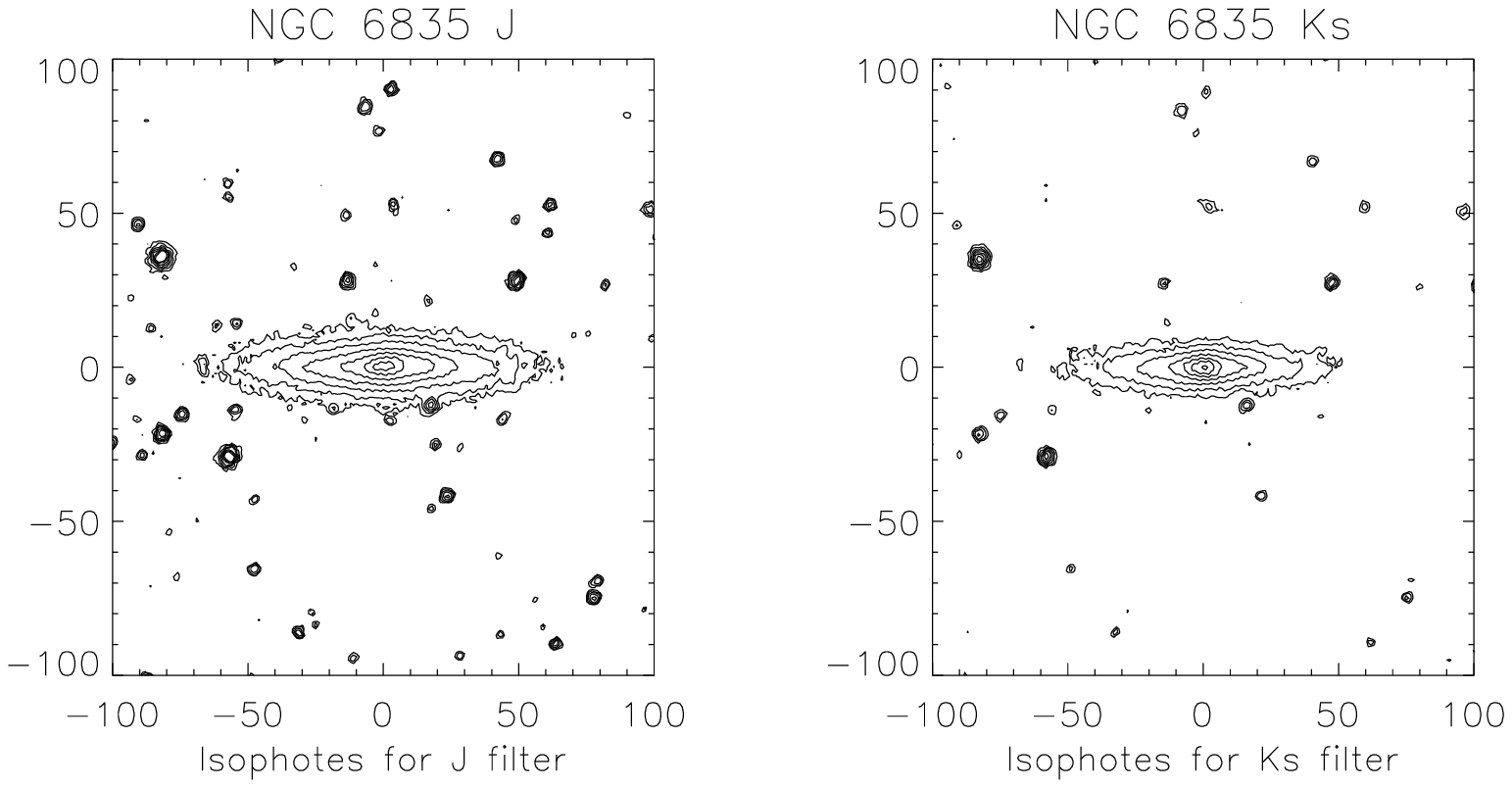}
\caption{continued}
\end{figure*}

\begin{figure*}
\setcounter{figure}{4}
\centering
\includegraphics[width=17cm]{3310f5b.epsi}
\caption{continued}
\end{figure*}
\begin{figure*}
\setcounter{figure}{4}
\centering
\includegraphics[width=17cm]{3310f5c.epsi}
\caption{continued}
\end{figure*}
\begin{figure*}
\setcounter{figure}{4}
\centering
\includegraphics[width=17cm]{3310f5d.epsi}
\caption{continued}
\end{figure*}
\begin{figure*}
\setcounter{figure}{4}
\centering
\includegraphics[width=17cm]{3310f5e.epsi}
\caption{continued}
\end{figure*}
\begin{figure*}
\setcounter{figure}{4}
\centering
\includegraphics[width=17cm]{3310f5f.epsi}
\caption{continued}
\end{figure*}
\begin{figure*}
\setcounter{figure}{4}
\centering
\includegraphics[width=17cm]{3310f5g.epsi}
\caption{continued}
\end{figure*}
\begin{figure*}
\setcounter{figure}{4}
\centering
\includegraphics[width=17cm]{3310f5h.epsi}
\caption{continued}
\end{figure*}
\begin{figure*}
\setcounter{figure}{4}
\centering
\includegraphics[width=17cm]{3310f5i.epsi}
\caption{continued}
\end{figure*}
\begin{figure*}
\setcounter{figure}{4}
\centering
\includegraphics[width=17cm]{3310f5j.epsi}
\caption{continued}
\end{figure*}
\begin{figure*}
\setcounter{figure}{4}
\centering
\includegraphics[width=17cm]{3310f5k.epsi}
\caption{continued}
\end{figure*}

{\bf Appendix A: z-integrating + deprojecting = deprojecting + z-integrating.}

\begin{figure}
\resizebox{\hsize}{!}{\includegraphics{3310f6.epsi}}
\caption{Definition of the involved quantities in Appendix A.}
\label{}
\end{figure}

For our purpose the two reduction steps of z-integrated and deprojecting can be interchanged, as we are only interested in obtaining the face-on view of the galaxy.

We observe $I(r,z)$, the surface brightness of the edge-on galaxy. We want to obtain $I(R)$, the surface brightness as a function of the galactocentric radius, for a face-on galaxy.

We take the pixel size as unity.

We can follow two procedures:

a) First summing in z, then deprojecting

b) First deprojecting, then summing in z.

In the first case, from $I(r,z)$ we obtain $I(r) = \Sigma_z I(r,z)$, then obtain $I(R)= {{I(r)} \over {A(R,r)}}$ (quantities are those defined in F01). This last equation is schematic. Of course, $I(R)$ is obtained by a series of more complicated steps but they are essentially of this type and to follow the series step by step does not introduce any change.

In the second case from $I(r,z)$ we obtain $l(R,z)$ as $l(R,z) = {{I(r,z)} \over {A(R,r)}}$ (observe that the areas $A$ depend on $R$ and $r$ but not on $z$). Then we sum in $z$
$$ 
   I(R) = \Sigma_z l(R,z)= \Sigma_z {{I(r,z)} \over {A(R,r)}}=
          {1 \over {A(R,r)}} \Sigma_z I(r,z) = {1 \over {A(R,r)}} I(r)
$$
obtaining exactly the same result as in the first procedure.

{\bf Appendix B: An inclination 90$^o >$i$>$88$^o$ does not introduce important errors.}

These errors depend not only on $i$ but also on the angular resolution. Suppose a galaxy with $i= 88^o$. Along the minor axis distances are reduced by a factor $\sin{2^o}= 0.035$. We want to know the size OO' perpendicular to the major axis (Fig. 7). In an exponential disk, the galaxy has no end and $\bar{OO'}= \infty$. However we want a characteristic value as at P the density is higher and decreases rapidly when we go from P to O'. We must find $I(x)$ for a given $r$.

We assume $I= I_o \exp{(-R/h)}$, hence $I(x) = I_o \exp{-(r^2+x^2)^{1/2}}$, where $r$ is a constant. First we work with unprojected distances.
By a Taylor expansion
$$
   (r^2+x^2)^{1/2} \approx r+ {1 \over 2} {x^2 \over r}
$$
therefore
$$
   I(x) = I_o e^{-r/h}e^{-{1 \over 2}{x^2 \over {rh}}}=
          I(P) e^{-{1 \over 2} {x^2 \over {rh}}}
$$
where $I(P)=I(R=r)$ is a constant.
\begin{figure}
\resizebox{\hsize}{!}{\includegraphics{3310f7.epsi}}
\caption{High inclination view of the galaxy.}
\label{}
\end{figure}
We now find a characteristic OO' distance, considering that it corresponds to a value of x at which $I(x)$ is $I(P)/e$
$$ e^{-{1\over 2} {x^2 \over {rh}}} = {1 \over e} $$
$$ {1 \over 2} {x^2 \over {rh}}=1$$
$$ x= \sqrt{2rh}$$.

\begin{figure}
\resizebox{\hsize}{!}{\includegraphics{3310f8.epsi}}
\caption{Definition of the involved quantities in Appendix B.}
\label{}
\end{figure}
For typical values in our galaxies, $r= 40''$, $h= 10''$, we have $x=30''$. And now we project $30'' \sin{2^o}= 1''$. This is of the order of the pixel size. Therefore our pixel size is too large to appreciate the $i \ne 90^o$ inclination. Therefore, variations in $z$ are mainly attributable to real variations in the vertical direction for $i \ge 88^o$.

\end{document}